\documentclass[twoside]{ilcws08}
\usepackage[latin1]{inputenc}
\usepackage[dvips]{graphicx,epsfig,color}
\usepackage{wrapfig,rotating}
\usepackage{amssymb,amsmath,array}

\begin{document}
\title{The {\tt POWHEG} method applied to top pair production and decays at the ILC
} 
\author{Oluseyi Latunde-Dada
\thanks{Cavendish--HEP--08/18}
\vspace{.3cm}\\
Cavendish Laboratory, University of Cambridge, \\
JJ Thomson Avenue, Cambridge CB3 0HE, U.K.
}

\maketitle

\begin{abstract}
We study the effects of gluon radiation in top pair production and their decays for
  $e^+e^-$ annihilation at the ILC. To achieve this we apply the {\tt POWHEG} method and
  interface our results to the Monte Carlo event generator \textsf{Herwig++}. We consider a center-of-mass energy of $\sqrt{s}=500$
  GeV and compare decay correlations and bottom quark distributions before
  hadronization.
\end{abstract}

\section{Introduction}
In Table \ref{tab:limits}, we have highlighted some differences between matrix element (ME) and
parton shower (PS) generators and have labelled as (M) or (D) those attributes we
consider merits or drawbacks respectively.
\begin{table}[!ht]
\centerline{\begin{tabular}{|l|r|}
\hline
\hline
{\bf PS generators}& {\bf ME generators} \\
\hline
\hline
Resums leading logarithmic contributions &   Can only go up to N$^{\sim
6}$ LO (D) \\
to all orders (M) & \\
\hline
\\
High multiplicity hadrons & Low multiplicity partonic \\
in the final state (M)  & final states (D)  \\
\hline
\\
Works well in regions & Works well in regions \\
of low relative $p_T$ (M \& D) & of high relative $p_T$ (M \& D)\\
\hline
\\
 Total rate is accurate to LO (D) &  Total rate is accurate to N$^{(>0)}$LO (M) \\
\hline
\hline
\hline
\end{tabular}}
\label{tab:limits}
\caption{Differences between PS and ME generators}
\end{table}
Note that most PS generators attempt to include NLO corrections via a method called
the matrix element correction which corrects the hardest shower emission {\it so far} to the exact matrix element and
populates the high $p_T$ regions according to the NLO cross-section. However, the total
rate is still only accurate to LO and virtual corrections are not fully taken care of.
\subsection{Getting the best of both worlds at NLO}
\begin{figure}[!ht]
\centerline{\includegraphics[width=0.45\columnwidth]{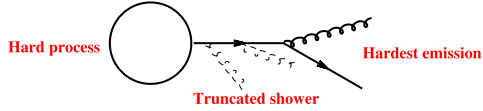}}
\label{schem}
\caption{POWHEG emissions}
\end{figure}

The Positive Weighted
Hardest Emission Generation ({\tt POWHEG}) method \cite{Nason:2004rx,Frixione:2007vw} aims
to solve this problem. It
\begin{enumerate}
\item generates total rates accurate to NLO,
\item treats hard emissions as in ME generators,
\item treats soft and collinear emissions as in PS generators,
\item and generates a set of fully exclusive events which can be interfaced with a
    hadronization model.
\end{enumerate} 
The {\tt POWHEG} method achieves this by generating the hardest emission in the shower first to NLO accuracy
using a modified Sudakov form factor. For angular ordered showers like {\tt Herwig++}, it
also includes a truncated shower of
    soft, wide angled emissions from the hard scale to the scale of the hardest
    emission. This maintains the correct soft emission pattern.
 This is illustrated in Figure \ref{schem}.
It then showers the resulting partons subject to a $p_{\rm T}$ veto to ensure that no harder
    emissions are generated. Unlike {\tt MC@NLO} \cite{Frixione:2002ik}, it is independent
    of the PS generator used and all events have positive weight.
In this talk \cite{url}, we will focus on the description and applications
    of the method in conjunction with the PS generator, {\tt Herwig++} \cite{Bahr:2008pv}. 



\subsection{The parton shower hardest emission cross-section}
For a single parton, the cross-section for the hardest emission with transverse
    momentum $p_{\rm T}$ is given by
\begin{equation}
\label{one}
d \sigma = d \sigma_{\rm B} \left[\Delta_{\rm V}(0)+\Delta_{\rm V}(p_{\rm T})\frac{\alpha_{\rm
      S}}{2 \pi} P dz\frac{dq^2}{q^2} \right] \,,
\end{equation}
where $P$ is the splitting function for the hardest emission and $\Delta_{\rm V}(p_{\rm T})$ is the Sudakov form factor for no emissions with $k_{\rm
  T} > p_{\rm T}$ which is given by
\begin{equation}
\Delta_{\rm V}(p_{\rm T}) = \exp \left[-\int dz \frac{dq^2}{q^2} \frac{\alpha_{\rm
      S}}{2 \pi} P \Theta(k_{\rm
  T} - p_{\rm T})\right] \;.
\end{equation}
The cross-section (\ref{one}) expanded to order $\alpha_{\rm S}$ gives
\begin{equation}
\label{three}
d \sigma = d \sigma_{\rm B} \left[\left\{1-\int \frac{\alpha_S}{2 \pi} P dz \frac{dq^2}{q^2}
  \right \} +\frac{\alpha_{\rm
      S}}{2 \pi} Pdz\frac{dq^2}{q^2} \right] \;.
\end{equation}
The {\tt POWHEG} method aims to substitute (\ref{three}) with the exact NLO result
    within the parton shower.
\subsection{Correcting to the exact NLO cross-section}
The exact NLO cross-section can be written as 
\begin{equation}
d \sigma_{\rm NLO} = d \sigma_{\rm B}+d \sigma_{\rm V}+d \sigma_{\rm R} \equiv  d \sigma_{\rm B}+d \sigma_{\rm V}+d
\sigma_{\rm B} dr {\mathcal M}  \;.
\end{equation}
Adding and subtracting $d \sigma _{\rm B} \int_{\delta} dr ({\mathcal M} - {\mathcal C})$ we get
\begin{eqnarray}
d \sigma_{\rm NLO} &=& d \sigma_{\rm B}+d \sigma_{\rm V} + d \sigma _{\rm B} \int_{\delta} dr
({\mathcal M} - {\mathcal C}) + d
\sigma_{\rm B} dr {\mathcal M} - d \sigma _{\rm B} \int_{\delta} dr ({\mathcal M} -
{\mathcal C}) \;.   
\end{eqnarray}
where ${\mathcal C}$ is a counter-term and $\delta$ is the subtraction region.
This can be rearranged to give 
\begin{eqnarray}
\label{six}
d \sigma_{\rm NLO}&=&d \sigma_{\rm V'} + d \sigma_{\rm B} \int_{\delta} dr ({\mathcal M} -
{\mathcal C})  + d \sigma_{\rm B} \left[\left \{1-\int_{\delta}dr {\mathcal M}\right \} + {\mathcal M} dr \right]
\end{eqnarray}
with $d \sigma_{\rm V'} = d \sigma_{\rm V}+d \sigma_{\rm B} \int_{\delta} dr {\mathcal C}$ now finite.
Comparing (\ref{six}) with (\ref{three}) above, we can write down an analog of (\ref{one}) as 
\begin{equation}
d \sigma_{\rm NLO} = d \sigma_{\bar{\rm B}} \left[\Delta_{\rm NLO}(0)+\Delta_{\rm
    NLO}(p_{\rm T}) \mathcal{M}dr \right]
\end{equation}
where
\begin{eqnarray}
d \sigma_{\bar{\rm B}} = d \sigma_{\rm B} + d \sigma_{\rm V'} + d \sigma_{\rm B} \int_{\delta} dr ({\mathcal M} -
{\mathcal C}) \nonumber \\
\Delta_{\rm NLO}(p_{\rm T}) =  \exp \left[-\int {\mathcal M} \Theta(k_{\rm
  T} - p_{\rm T})\right] \;.
\end{eqnarray}
Note that in defining $d \sigma_{\bar{\rm B}}$, we have neglected terms of higher order than
$\alpha_{\rm S}$ and if it is negative, perturbation theory has broken down.
\subsection{{\tt POWHEG} formalism and applications}
With 
 \begin{equation}
d \sigma_{\rm NLO} = d \sigma_{\bar{\rm B}} \left[\Delta_{\rm NLO}(0)+\Delta_{\rm
    NLO}(p_{\rm T}) \mathcal{M} dr\right] \,,
\end{equation} 
the {\tt POWHEG} method can be applied by
\begin{enumerate}
\item generating the $p_{\rm T}$ of the hardest
emission and its emission variables $r$, according to the term in square brackets using well known
Monte Carlo techniques,
\item distributing the underlying Born variables according to $d \sigma_{\bar{\rm
        B}}$, (This defines the event weight and since it is always positive definite, all
    event weights are positive)
\item for angular ordered showers, implementing a truncated shower of soft emissions
    between the {\it hard} scale and the scale of the hardest emission,
\item and finally showering the resulting partons as in a PS
  generator subject to a $p_{\rm T}$ veto.
\end{enumerate}

The {\tt POWHEG} method has been applied successfully to the following
processes $Z$ pair hadroproduction \cite{Nason:2004hfa}, heavy flavour production
\cite{Frixione:2007nw}, $e^+e^-$
annihilation to hadrons \cite{Latunde-Dada:2006gx}, Drell-Yan vector boson production \cite{Alioli:2008gx,Hamilton:2008pd} and Higgs boson
production via gluon fusion \cite{Alioli:2008tz}. 
\section{Top-pair production and decay at the ILC}
\begin{figure}
\centerline{\includegraphics[width=2.5in, height =2in]{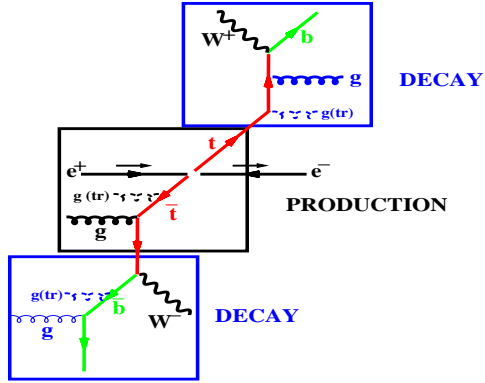}}
\label{Diag}
\caption{Top-pair production and decay}
\end{figure}
The application to top pair production and decay at the ILC takes the following into
account.
\begin{enumerate}
\item Spin correlations are taken into account in the
    matrix elements, ${\mathcal M}$ for the 
production and decays of the top pairs.
\item The narrow width approximation is applied so that production and decay interference
    can be neglected. This independence enables us to apply the method in separate
    frames: the lab frame for production and the top rest frame for its decay.
\item In the lab frame, the transverse momentum $k_T$ is defined relative to the original
    $t-\bar{t}$ axis whilst in the top rest frame it is relative to the original $b-W$ axis.
\item The scale range available for production emissions  ($\approx \log (\sqrt{s}/m_{\rm
        t})$) and is much less than the range available for decay emissions ($\approx \log
      (m_{\rm t}/m_{\rm
        b})$).
\item There are two different sources of the decay emissions: one from the top quark
    before it decays and the other from the $b$ quark after the decay. Hence, there are
    three different regions for truncated emissions labelled {\bf g(tr)} in Figure \ref{Diag}. These are before the hardest emission in the production, from the
  top quark before it decays and before the hardest emission from the $b$
  quark.\end{enumerate}
Further discussion of the method and its application can be found
    in \cite{Latunde-Dada:2008bv}. 
Setting $\sqrt{s}=500$ GeV and $m_{\rm t}=175$ GeV, we considered the following four cases
with {\tt POWHEG} interfaced with {\tt Herwig++}:
\begin{enumerate}
\item Leading order (LO) with no {\tt POWHEG} emissions,
\item Only {\tt POWHEG} emissions in the production process (Prod),
\item Only {\tt POWHEG} emissions in the decays of the tops (Dec),
\item Both production and decay emissions allowed (Prod + Dec).
\end{enumerate}
For the two different $e^{+}e^{-}$ initial polarizations, we investigated the correlations
between the decay products (we consider leptonic decays only for the $W$ bosons) and the momentum
distributions of the $b$ quarks before hadronization. A selection of the plots obtained
are presented in Figure 3.
\begin{figure}[!ht]
\hspace{0cm}
\psfig{figure=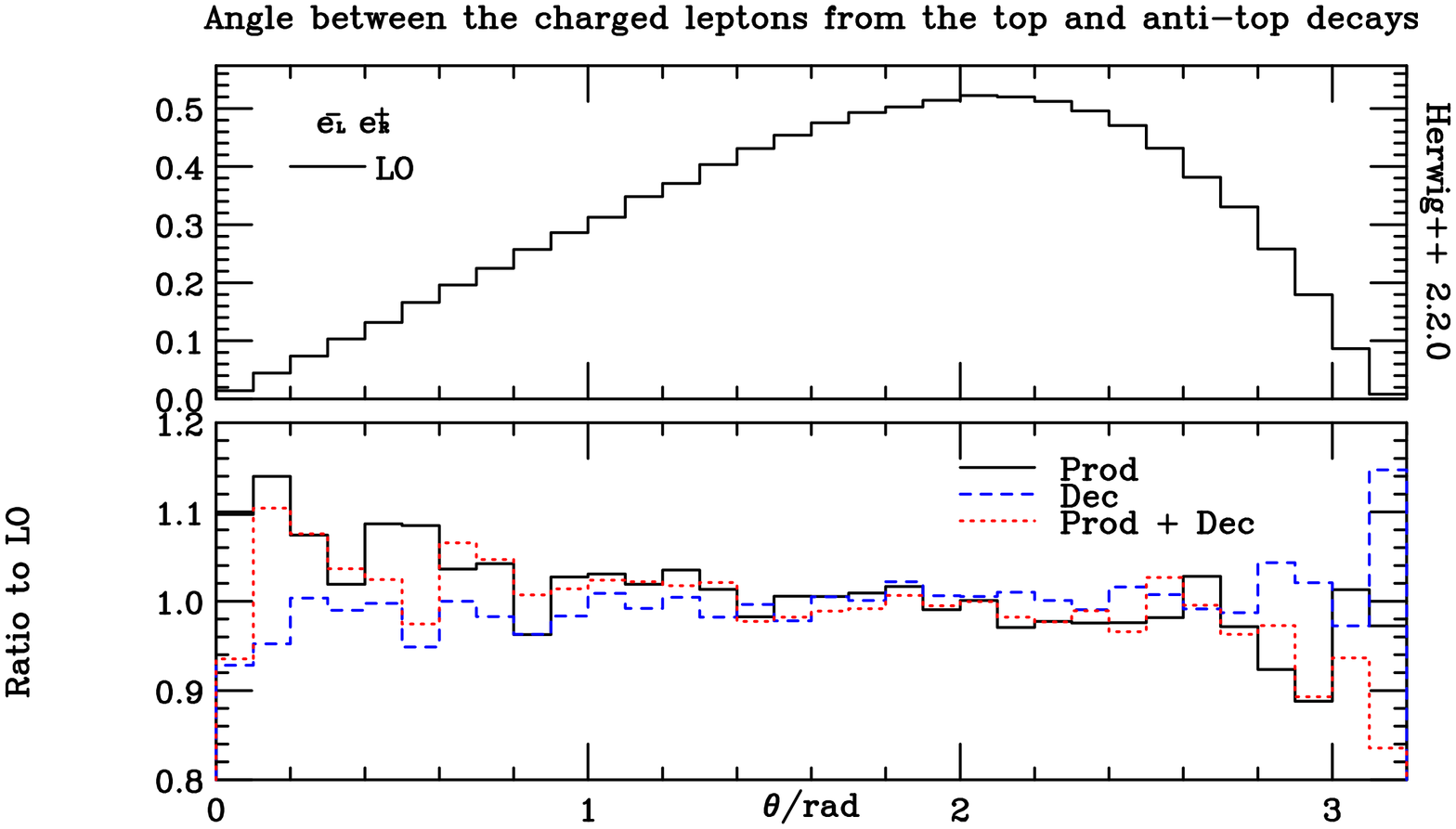,%
width=1.6in,height=3in,angle=90}
\psfig{figure=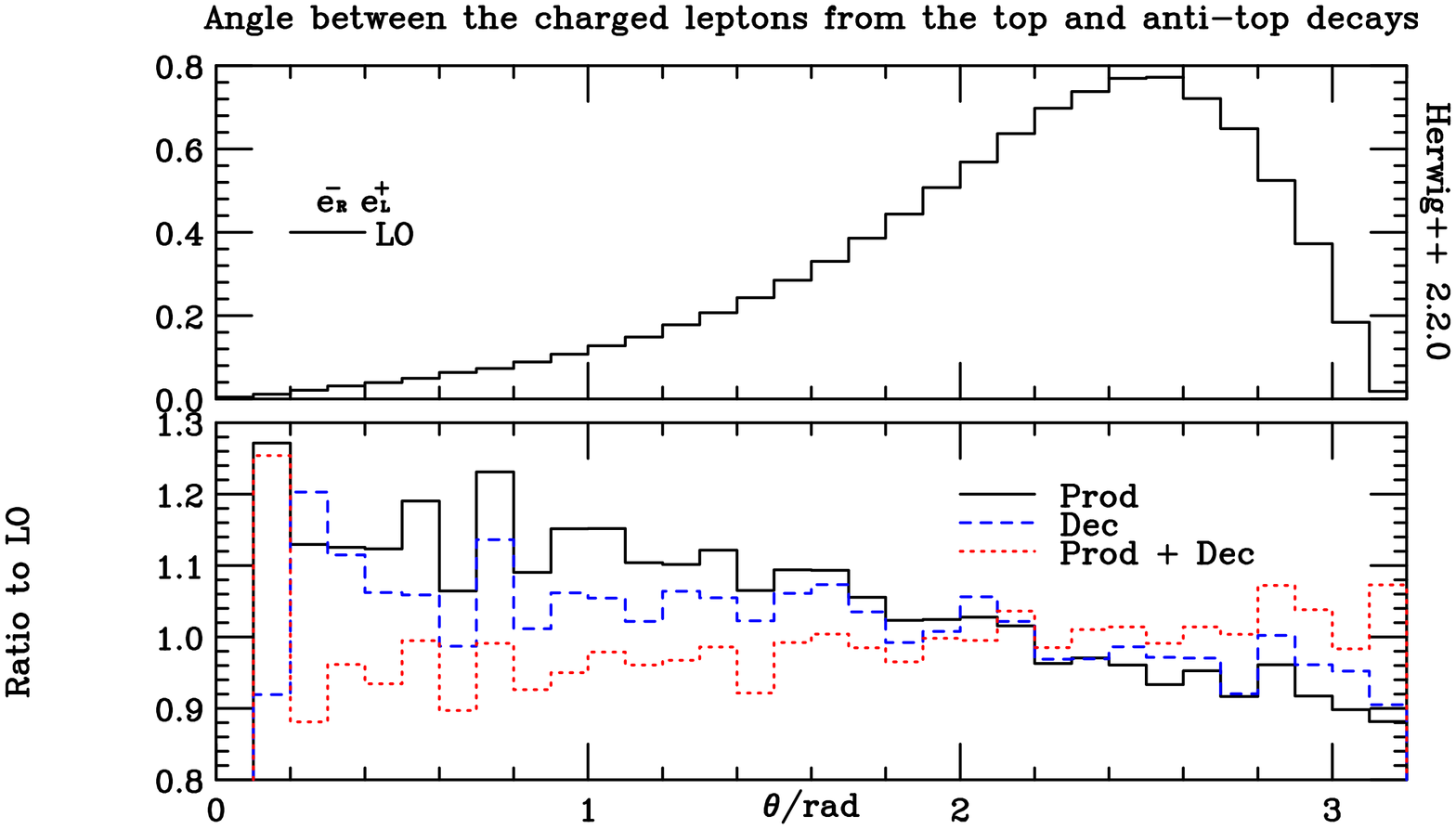,%
width=1.6in,height=3in,angle=90}
\end{figure}   
\begin{figure}[!ht]
\vspace{0cm}
\hspace{0cm}
\psfig{figure=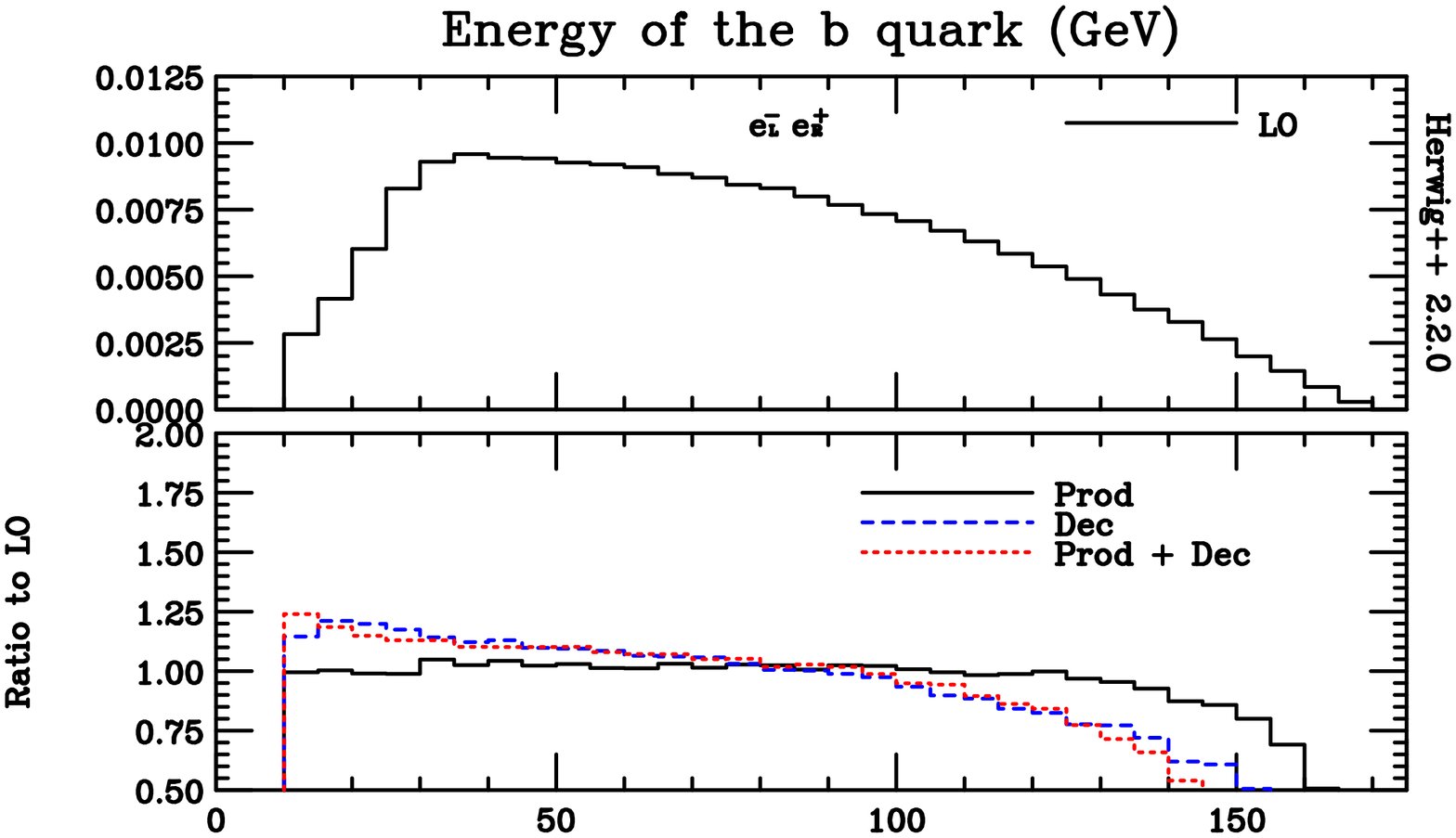,%
width=1.6in,height=3in,angle=90}
\psfig{figure=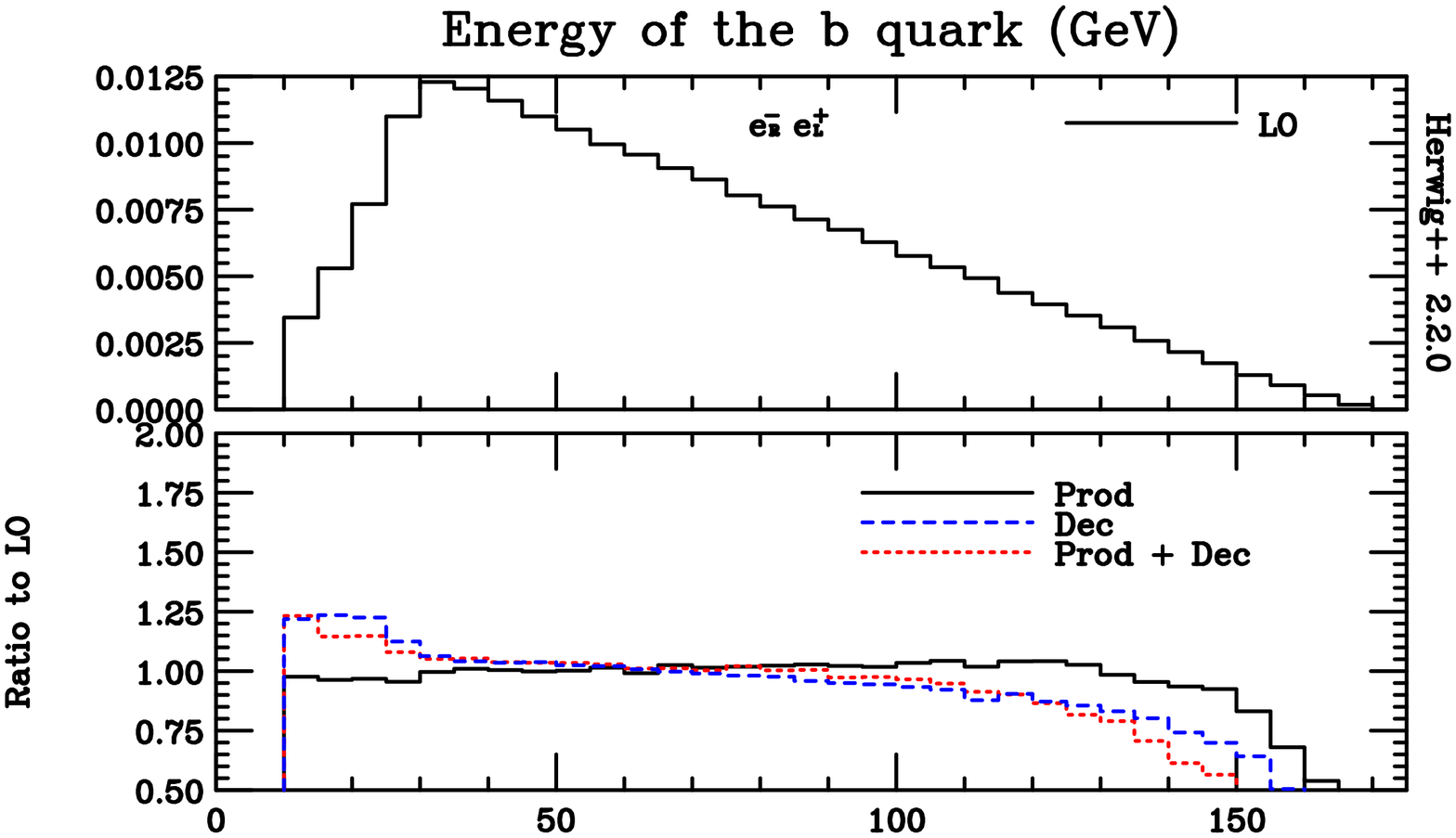,%
width=1.6in,height=3in,angle=90}
\end{figure}   
\begin{figure}[!ht]
\vspace{0cm}
\hspace{0cm}
\psfig{figure=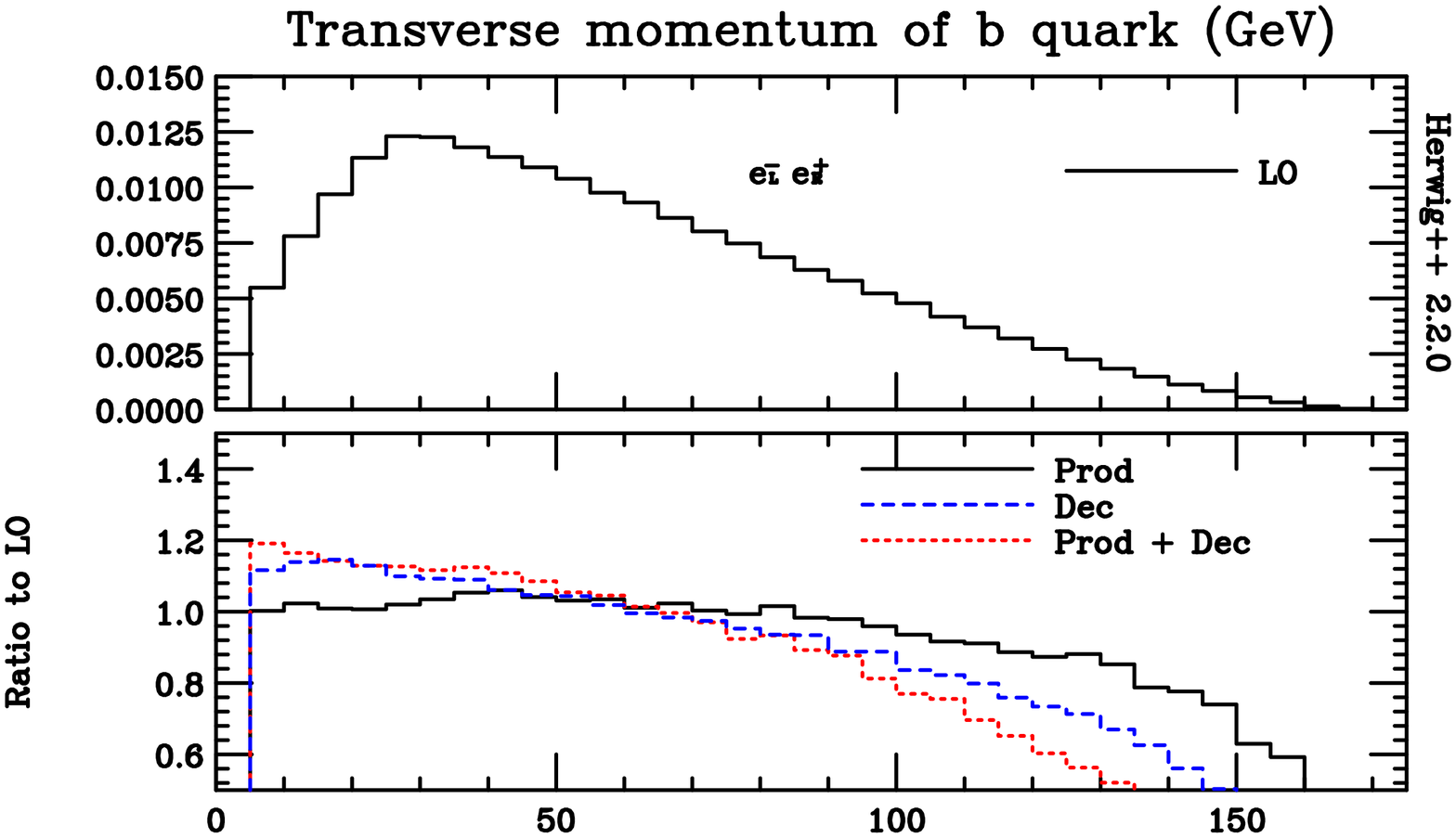,%
width=1.6in,height=3in,angle=90}
\psfig{figure=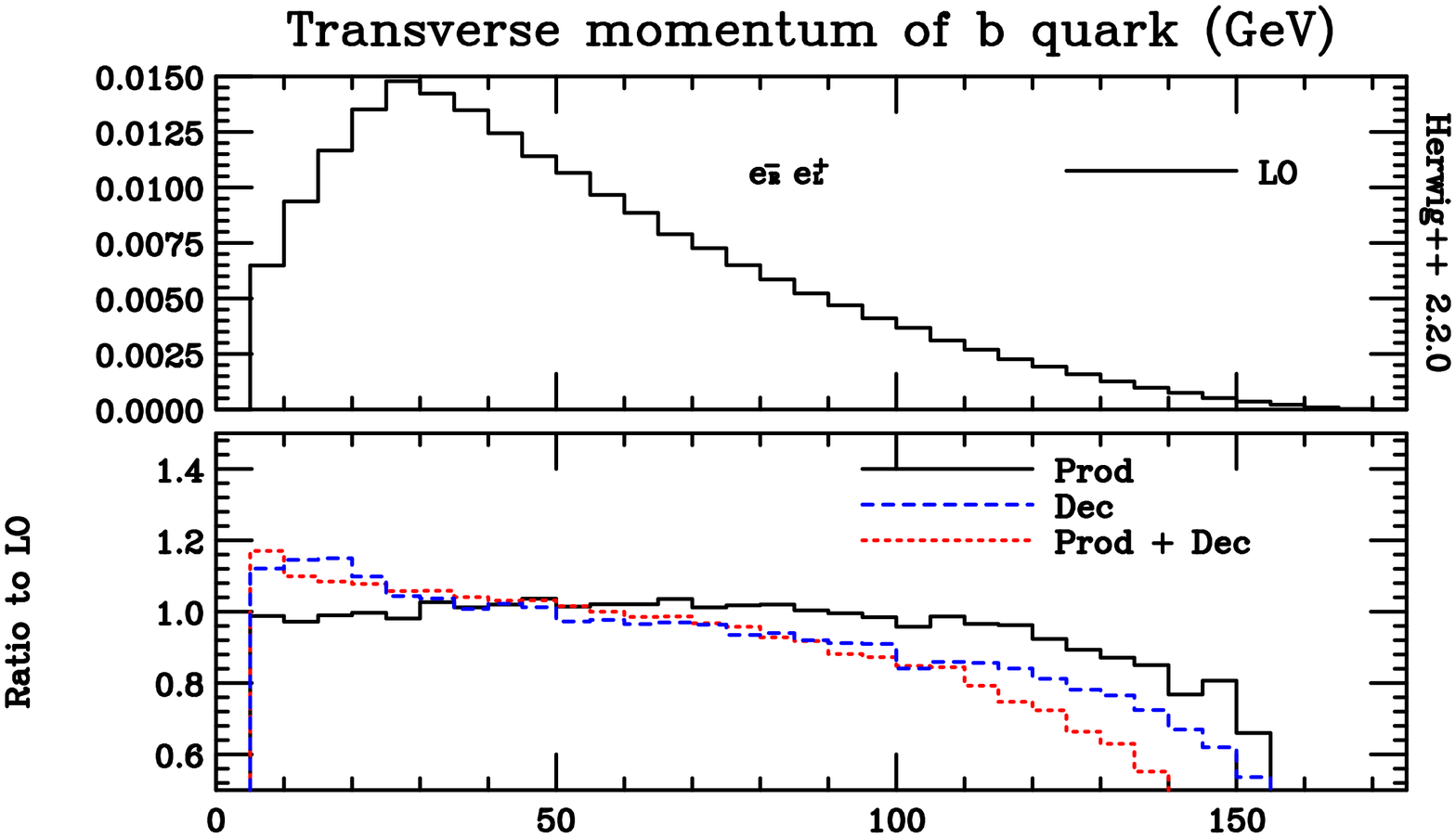,%
width=1.6in,height=3in,angle=90}
\end{figure}   
\begin{figure}[!ht]
\vspace{0cm}
\hspace{0cm}
\psfig{figure=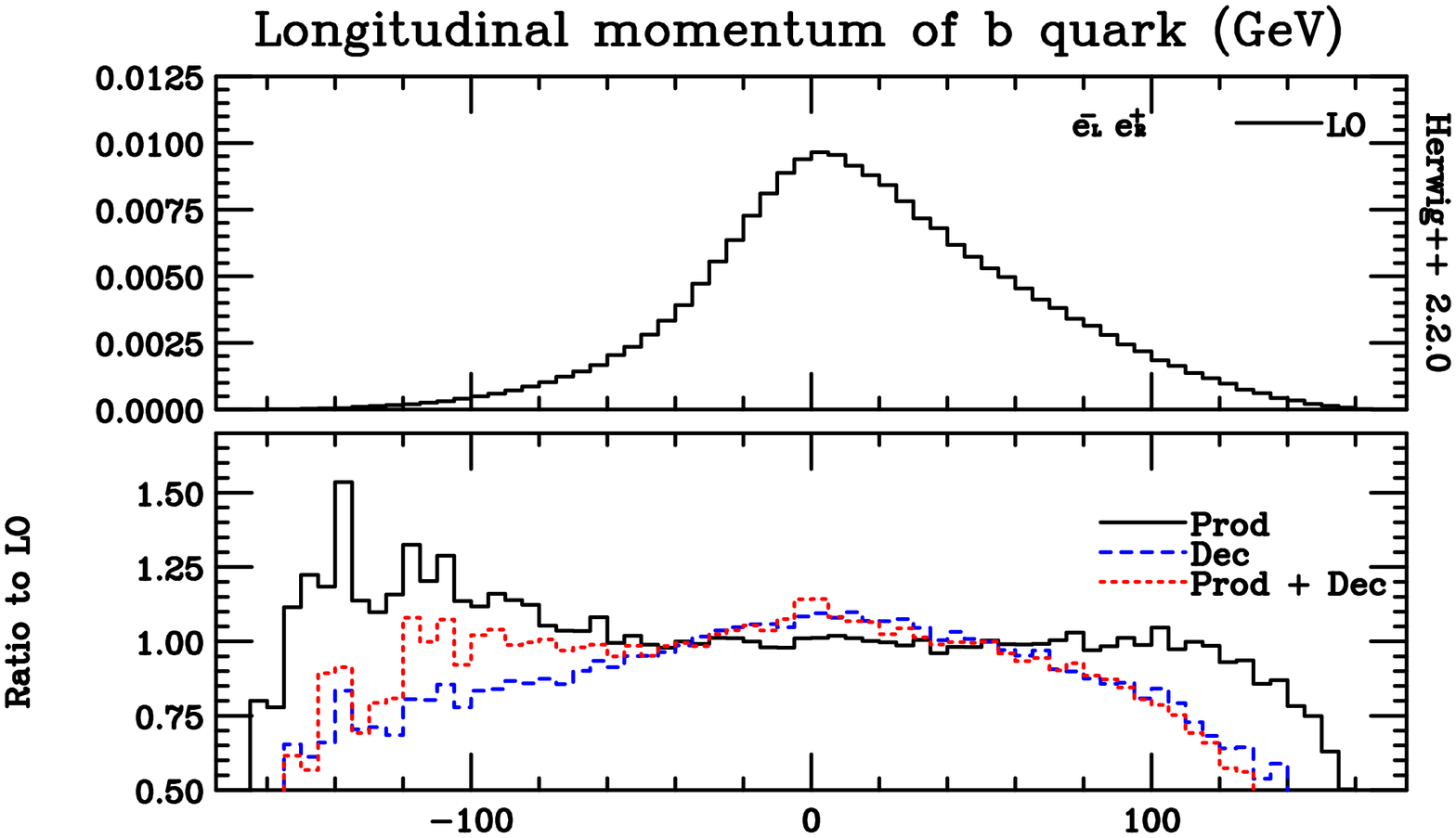,%
width=1.6in,height=3in,angle=90}
\psfig{figure=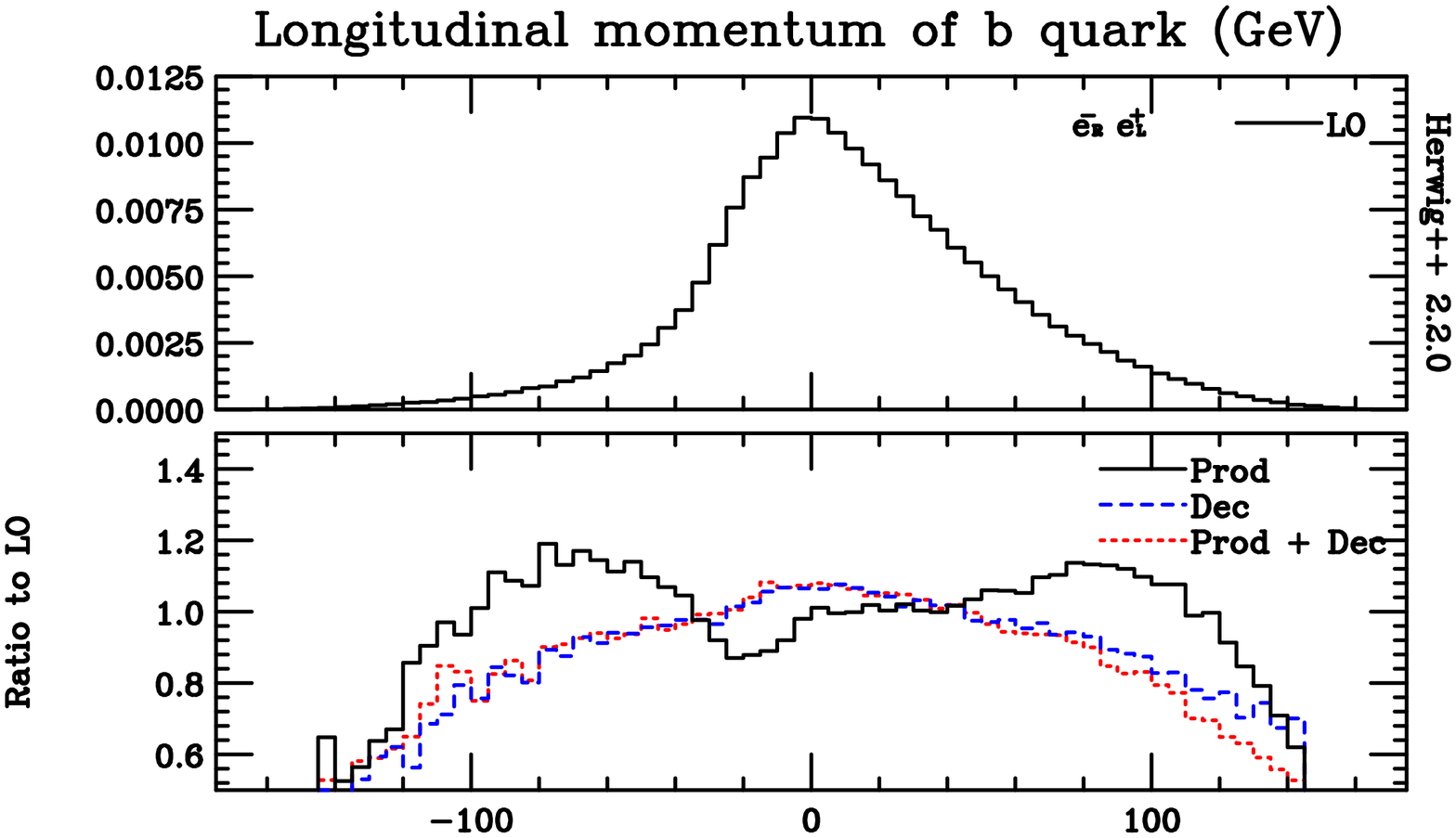,%
width=1.6in,height=3in,angle=90}
\label{plots}
\caption{Correlations and $b$ quark momentum distributions.}
\end{figure}



\section{Summary}

NLO improvements of parton showers are essential for near accurate predictions of angular
correlations and momentum distributions at future colliders. The {\tt POWHEG} method achieves this by distributing the {\it hardest} emission according
  to the NLO matrix element and yields events with positive weights. For angular ordered
  showers, the addition of a truncated shower is required. The method, though not very straightforward to apply, has demonstrated success in
    comparison with existing collider data. 

In this talk, we have extended this to top-pair production and
    decay at the ILC and made predictions for some distributions in comparison to leading
    order predictions. 

\section*{Acknowledgments}
We are grateful to the other members of the \textsf{Herwig++} collaboration for
developing the program that underlies the present work and for helpful comments. We are particularly grateful to
Bryan Webber for constructive comments and discussions throughout. This research was
supported by the Science and Technology Facilities Council, formerly the Particle Physics
and Astronomy Research Council, and the European Union Marie Curie Research Training
Network MCnet, under contract MRTN-CT-2006-035606.








\begin{footnotesize}



%

\end{footnotesize}


\end{document}